# Black-box Adversarial Transferability: An Empirical Study in Cybersecurity Perspective

Khushnaseeb Roshan*,a, Aasim Zafara,b

*aDepartment of Computer Science, Aligarh Muslim University, Aligarh, Uttar Pradesh, India*
*\*Email:kroshan@myamu.ac.in*
*bEmail:azafar.cs@amu.ac.in*

**Abstract**

The rapid advancement of artificial intelligence within the realm of cybersecurity raises significant security concerns. The vulnerability of deep learning models in adversarial attacks is one of the major issues. In adversarial machine learning, malicious users try to fool the deep learning model by inserting adversarial perturbation inputs into the model during its training or testing phase. Subsequently, it reduces the model confidence score and results in incorrect classifications. The novel key contribution of the research is to empirically test the black-box adversarial transferability phenomena in cyber attack detection systems. It indicates that the adversarial perturbation input generated through the surrogate model has a similar impact on the target model in producing the incorrect classification. To empirically validate this phenomenon, surrogate and target models are used. The adversarial perturbation inputs are generated based on the surrogate-model for which the hacker has complete information. Based on these adversarial perturbation inputs, both surrogate and target models are evaluated during the inference phase. We have done extensive experimentation over the CICDDoS-2019 dataset, and the results are classified in terms of various performance metrics like accuracy, precision, recall and f1-score. The findings indicate that any deep learning model is highly susceptible to adversarial attacks, even if the attacker does not have access to the internal details of the target model. The results also indicate that white-box adversarial attacks have a severe impact compared to black-box adversarial attacks. There is a need to investigate and explore adversarial defence techniques to increase the robustness of the deep learning models against adversarial attacks.

## 1. Introduction

The rapid developments in the field of Artificial Intelligence (AI) with such widespread application across multiple domains like cyber security [1][2][3], image classification [4][5][6], healthcare [7][8][9], and much more, giving rise to security concerns and robustness of machine learning (ML) and deep learning (DL) based applications. Recently, it has been observed that ML and DL models are susceptible to adversarial attacks in which hackers try to deceive the ML/DL model by inserting fabricated adversarial perturbation inputs either in the training or testing phase. For example, in cyber security, adversarial attacks aim to exploit vulnerabilities in DL-based Network Intrusion Detection Systems (NIDS). These attacks involve manipulating input traffic data by adding tiny adversarial noise and evading intrusion detection during the inference phase of NIDS. Similarly, in image classification, the adversary aims to mislead the model by classifying an image into the wrong category with high confidence. This is a serious security threat to the reliability of DL-based image classification models, as adversaries can manipulate input images to produce incorrect classifications. In healthcare systems, for example, during the COVID-19 pandemic, adversarial attacks could have significant consequences. For instance, a person not wearing a mask might be classified as wearing one, or vice versa, due to adversarial manipulations. Such misclassifications can have severe implications for public health and safety.

These examples underscore the critical need for research and development in adversarial machine learning to enhance the robustness and security of AI applications across diverse domains. As AI continues to play a pivotal role in shaping various aspects of our lives, addressing these security concerns becomes imperative to ensure the reliability and trustworthiness of AI-based systems. It is critical to ensure the safety of ML and DL systems in the real physical world. Furthermore, much research has been done to improve the performance of ML and DL systems and optimize their performance metrics [10] [11]. However, the generalization and robustness capability can not be ignored in today's era as unknown cyber security threats emerge daily [12].

Adversarial machine learning combines the fields of security and ML [13]. From a cybersecurity perspective, the robustness and security of the cyber attack detection system should not be compromised. Cyber attack detection systems (e.g., intrusion detection systems, intrusion prevention systems) are vulnerable to adversarial attacks. Hackers can easily fool any ML and DL cyber attack detection system by adding a tiny perceptible perturbation, which can lead to a malfunctioning model with incorrect classification results. There are two broad categories of adversarial machine learning, namely, white-box and black-box. In white-box adversarial attacks, the malicious users have complete access to the hyper-parameters of the target system, such as its gradients and model architecture. However, in the black-box adversarial attack, the malicious users have no or little information about the target system. The other categories are targeted and untargeted adversarial attacks. The malfunction model can produce the incorrect specific target label in the targeted attack. For example, the cyber attack detection system classifies the benign network traffic into any specific attack class and vice-versa. However, in untargeted attacks, the malicious user reduces only the confidence score (e.g., accuracy, f1-score, etc.) of the target system.

Szegedy et al. [14] highlighted the cross-data transferability property of the ML and DL models. It says adversarial perturbation input generated to fool one model can also trick another model trained on another or a different subset of the same dataset into producing incorrect output. In real-world situations, the malicious user does not have any information (e.g. hyperparameters, gradients, model architecture) of the target system but may have access to the training data. Hence, the hacker can use this phenomenon to produce a black-box attack in any ML and DL model. In this research work, we experimentally implemented the black-box transferability property in a cyber attack detection system. We have built two models. The first is the surrogate model which is used to create the adversarial perturbation examples. The second one is the target system that the hacker wants to exploit. Both models have different hyperparameters and internal architecture but are trained on the same dataset. Fast Gradient Sign Method (FGSM) [15], a famous technique, is used to generate adversarial perturbation.

Furthermore, the motivation and the need for the study lie in the exploration of the intriguing phenomena of DL models within cybersecurity [14], [15]. There is an absence of research work which explores this phenomenon specifically within the application of network security. Our investigation represents real-world and practical scenarios in which the adversaries have limited and no access to the target system. Through our study, we discovered that adversaries can successfully implement adversarial attacks even without access to the target system information, such as gradients and hyperparameters. In addition, this study is not only significant for network security but also holds implications for other domains. Our research contributes to the vulnerabilities of DL models against adversarial transferability black-box attack.

To the best of our knowledge, this research study is a novel contribution that practically implements the black-box adversarial transferability property from a cybersecurity perspective. A lot of research is available on adversarial attack and defence methods in multiple research areas, including cybersecurity [16] [17] [18] [19] [20] [21]. However, the major research gap is that none of the authors has empirically examined black-box adversarial attack transferability phenomena in cyber security. Hence, this research work would be a novel contribution to the network security domain and can guide network administrators to safeguard cyber attack detection systems against adversarial attacks.

The significant contributions of the research study are as follows:

- Implementation of the black-box untargeted adversarial transferability phenomena by using the concept of surrogate and target model in cyber attack detection system based on DL. The latest CICDDoS-2019 dataset is used for the experimentation.

- The adversarial perturbation examples are generated using the FGSM [15] and surrogate model with multiple epsilon values over the testing dataset.

- Both the target and surrogate model are evaluated with various performance metrics like accuracy, precision, recall and f1-score.

- Finally, the future scope is discussed, which will guide new researchers in finding a suitable direction in adversarial machine learning.

- We have used the simplistic approach and tried to explain every technical aspect of the proposed research work so that the new researcher can easily replicate this study for further enhancement.

The overall organization of the research study is as follows: the taxonomy of adversarial machine learning are discussed in Section 2. It includes a brief historical background followed by a broad categorization of adversarial attacks. Section 3 describes the related work. It highlights the most recent study, which combines the domain of cyber security and adversarial machine learning. Section 4 represents the description of the latest CICDDoS-2019 dataset used for the experimentation. Section 5 is the methodology that describes the complete overview and conceptual architecture of the proposed research work. It explains every technical aspect of the research work. Section 6 is the experimental and supportive library section. It also describes the testbed setting used for the implementation of the proposed research work. Section 7 is the results and discussion section in which both the models are evaluated before and after the adversarial attack. Section 8 is the future scope, followed by a conclusion in Section 9.

## 2. Adversarial Attack Taxonomy

In this section, we discussed the historical background, followed by the systematic categorization and terminology related to adversarial machine learning. We followed the simplistic approach for the reader's understanding; however, for more detail, the researcher can refer to the research articles [16] [17] [22]. Some survey and review research papers also provide a detailed background of adversarial machine learning and their key research areas [23] [24] [25].

The concept of adversarial machine learning appeared in the 2000s. However, in the last two decades, it has gained prominence due to growing interest in the security and robustness of ML and DL models. In 2004, Dalvi et al. [26] explored vulnerabilities in machine learning. The authors purposefully modified the email body to deceive the spam classifier. Later, in 2006, the author heightened the broad question, "Can machine learning be secure?" In 2010 [27], they studied the security of machine learning algorithms and suggested its defence strategies. A famous study [14] on the cifar-10 image dataset revealed the vulnerability of the image DL classification model. The author discovered susceptibility, indicating that adding a tiny amount of noise in the image can deceive the DL model to produce the wrong result with high confidence.

The four broad categorizations of adversarial machine learning are: 1) based on attack type, 2) based on timing, 3) based on model information, and 4) based on the purpose. The first category is further divided into four classes, namely, evasion, extraction, poisoning and inference. In an evasion adversarial attack, the malicious actor attempts to deceive the model by adding carefully crafted adversarial perturbation into the input data. For example, the cyber attack detection model classifies the benign instance into the attack class or the attack class into the benign data sample. In the extraction adversarial attack, the malicious actor attempts to steal the sensitive information of the target system, like its internal structure, hyperparameters, and training data, to make a substitute model that can mimic the behaviour of the target system. Later, this substitute model can be used to exploit the vulnerability of the target system. The third category is the poisoning attack, in which the corrupted input is inserted into the target model during its training phase. Poisoning attacks are more dangerous and have long-lasting effects because the model is trained on the corrupted data, which would produce incorrect classification. The fourth category is the inference attack. Unlike a poisoning attack, an inference attack attempts to fool the model during its decision-making or testing phase by inserting adversarial perturbation input and causing the model to produce an incorrect prediction.

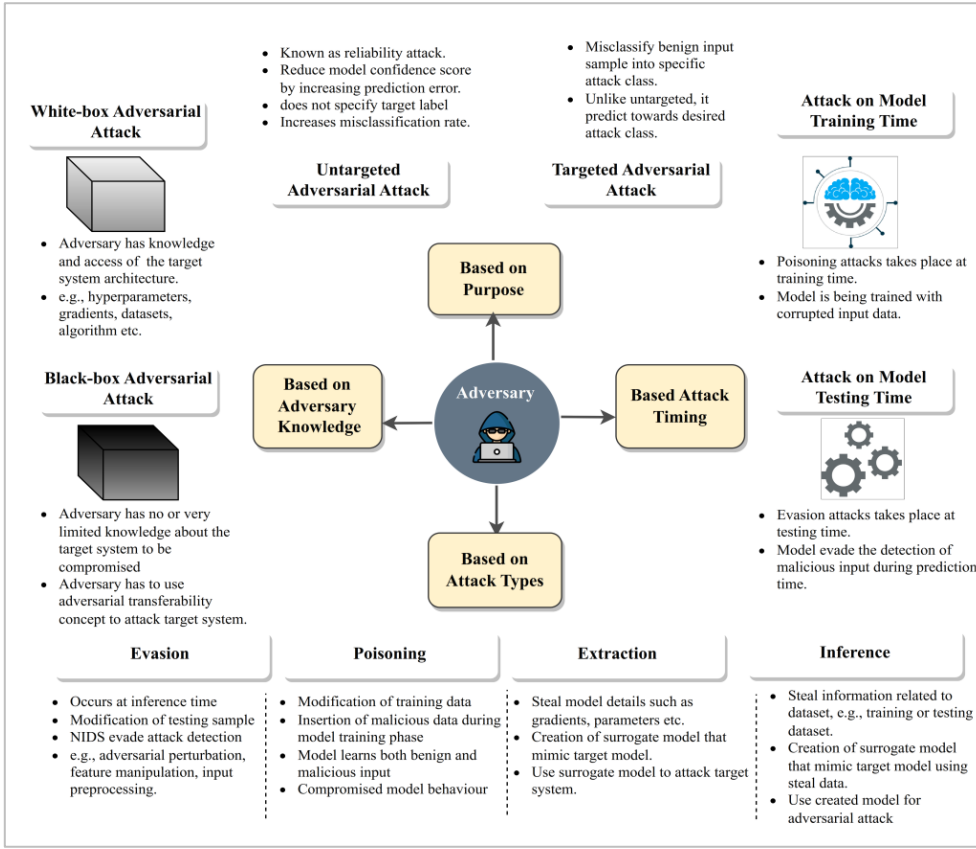

Figure 1 Adversarial Attack Taxonomy

Based on the timing, the adversarial attack can be encountered in the training phase or in the inference phase. Examples of this phenomenon are data poisoning attacks that try to deceive the model during the training phase. On the contrary, the evasion adversarial attacks aim to fool the model during testing. The third group is based on the model information that categorises black-box and white-box attacks. In a black-box attack, the attacker has no knowledge of the target model (like gradients, weights and internal parameters). The black-box attack represents more realistic scenarios because, in the real physical world, attackers possess very little knowledge of the target system they are trying to compromise. The last classification is based on the purpose of the attacker, which leads to the categorization between targeted and untargeted adversarial attacks. In the targeted attack, the main goal of the hacker is to craft the adversarial perturbation input in such a way that the model not only makes the wrong prediction but classifies it into the specific targeted label chosen by the hacker. On the contrary, in the untargeted attack, the goal is to disrupt the model decision-making process and reduce its confidence score in terms of classification metrics like accuracy and precision by increasing the prediction error.

All the dimensions of adversarial machine learning we discussed above are demographically presented in Figure 1 with a brief description. In the proposed research work, we have used the untargeted attack, using the white-box FGSM method during the inference or testing phase of the NIDS model. It is a type of evasion attack strategy where the NIDS model evades intrusion detection.

## 3. Related Work

In the last two decades, adversarial machine learning has become very popular. Many research article has been published in computer vision [28] [17] [29], but less is explored from the computer network security point of view [30] [31]. In this research work, we have mainly focused on adversarial machine learning from a computer network security perspective. Moreover, the generation of adversarial perturbation in constrained or structured domains (e.g. network traffic logs) is different compared to the image domain. In the image domain, the feature space is the pixel value. It is easy for the adversary to exploit each feature or pixel value of the image. On the contrary, in network traffic data logs, the feature value can be continuous, binary or categorical. The features may be correlated or may have fixed values that can not be modified by the adversary. The most related research work in network anomaly and intrusion detection is given in the following.

Clements et al. [32] conducted a research study into the robustness of the deep learning-based Network Intrusion Detection System (NIDS). The author used Kitsune [33], a lightweight ensemble model based on autoencoders designed for online network anomaly detection. It comprises components including packet capture, feature extraction, feature mapping, and anomaly detection. The main goal is to investigate the system against adversarial attacks. To assess its effectiveness, four different adversarial algorithms—namely, Fast Gradient Sign Method (FGSM), Jacobian-based Saliency Map Attack (JSMA), Carlini and Wagner (C&W), and Expectation over Normalized Mean (ENM)—were employed. Notably, the attacker had the knowledge of the target model and was able to directly generate adversarial perturbed inputs.

Usama et al. [34] used Generative Adversarial Networks (GANs) to create adversarial examples to evade NIDS. This research demonstrated that GANs could effectively counter adversarial perturbations, thus enhancing the model's resilience against adversarial attacks. The author employed a unique method to alter only non-functional features of the network traffic data during both the attack and defence phases. This strategy aimed to ensure that the adversarial manipulations did not significantly impact the network's genuine operations. The proposed work is evaluated using the KDDCUP-99 benchmark dataset with multiple ML algorithms, including Deep Neural Networks (DNN), Support Vector Machines (SVM), Decision Trees (DT), Random Forest (RF), and others. The results are evaluated with key metrics like accuracy, precision, recall, and f1-score under four scenarios: before the attack, after the attack, after the adversarial defence, and after the GAN-based adversarial defence.

Pawlicki et al. [35] have done an in-depth exploration and analysis of the impact of adversarial attacks, specifically FGSM, Carlini and Wagner (C&W), Projected Gradient Descent (PGD), and Basic Iterative Method (BIM), across various ML classifiers. These classifiers include Artificial Neural Networks (ANN), RandomForest (RF), AdaBoost, and Support Vector Machine (SVM)., all within the domain of network security. The study aimed to mitigate the vulnerabilities exposed by adversarial perturbations and was conducted over the CICIDS-2017 dataset.

A similar study has been conducted by Guo et al. [36] conducted a comprehensive analysis of the impact of the Basic Iterative Method (BIM), a type of black-box adversarial attack, across five different Machine Learning classifiers. The study encompassed Convolutional Neural Networks (CNN), Support Vector Machine (SVM), k-Nearest Neighbor (KNN), Multilayer Perceptron (MLP), and Residual Network (Resnet). The two benchmark intrusion detection datasets, KDDCUP-99 and CICIDS-2017, are used to experiment and validate the results. The evaluation aimed to assess the susceptibility of these classifiers to adversarial manipulations. However, the author suggested two future studies; the first is applying it to real network traffic datasets. And second is to explore another complex adversarial attack.

Alhajjar et al. [37] also explored the same area but with an evolutionary approach. The authors employed advanced techniques to generate adversarial examples. These include the purpose of deceiving Machine Learning (ML) and Deep Learning (DL) models. Their research integrated three distinct methods: Particle Swarm Optimization (PSO), Genetic Algorithm (an evolutionary computation technique), and Generative Adversarial Network (GAN). These techniques were harnessed to create adversarial examples with the intention of misleading both ML and DL models. The experimentation is carried out over two datasets, namely, UNSW-NB15 and NSL-KDD.

Sethi et al. [38], the authors proposed a context-adaptive intrusion detection system (IDS). It leveraged distributed deep reinforcement learning agents across the network. Through extensive experimentation on NSL-KDD, UNSW-NB15, and AWID datasets, the proposed model outperforms existing systems in terms of accuracy and false positive rates. The IDS demonstrates resilience against adversarial attacks and enhances its robustness to mitigate its effects. The proposed method used denoising autoencoder into the system.

Zhang et al. [39], the authors proposed three DNN architectural CNN, and LSTM for NIDS models. The results obtained on the CSE-CIC-IDS2018 dataset yield models with a good 98.7% detection accuracy. To demonstrate evasion of adversarial attack, the authors employ five advanced attack strategies—NES, BOUNDARY, HOPSKIPJUMPATTACK, POINTWISE, and OPT-ATTACK. Adversarial samples manipulate traffic features within realistic domain constraints for model evaluation purposes. The proposed study is novel in terms of adversarial attacks, as most of the researchers studied only FGSM, PGD, and JSMA gradient-based attacks.

Han et al. [40], the research introduced an exploration of grey/black-box adversarial attacks on ML-based NIDS. The proposed automatic attack achieves >97% evasion rate in half the cases for Kitsune, a state-of-the-art NIDS. The study also proposed a defence method that reduced evasion rates by >50% in most cases. These findings provide critical insights into addressing adversarial challenges in ML-based NIDSs.

Maarouf et al. [41], explored the resilience of ML and DL methods in classifying encrypted internet traffic under adversarial attacks. The proposed research used C4.5 Decision Tree, K-Nearest Neighbor (KNN), Artificial Neural Network (ANN), Convolutional Neural Networks (CNN), and Recurrent Neural Networks (RNN), DL based models. The study evaluates the performance on two benchmark datasets, ISCX VPN-NonVPN and NIMS, using Mutual Information for feature selection Notably, deep learning exhibits better resilience against adversarial samples compared to machine learning in most experimental results. The author also assessed the effectiveness of three adversarial attacks – Zeroth Order Optimization (ZOO), Projected Gradient Descent (PGD), and DeepFool. The comparative analysis of DL and ML in both adversarial-free and adversarial attack environments provides valuable insights into the classification of encrypted traffic in terms of classification reports.

Sarıkaya et al. [42], introduced a novel approach to combat adversarial attacks on machine learning-based intrusion detection systems (IDS). The method incorporated generative adversarial networks (GANs) for the generation of adversarial attack data, demonstrating the effectiveness of GAN-based attacks on ML-based IDS/IPS systems. Key contributions involved utilizing the reconstruction error values from autoencoders as inputs for detecting adversarial examples. Additionally, a LightGBM classifier was trained based on the predictions generated by these autoencoders. The study aimed to develop a generalized robust IDS model effective against various common adversarial attack types, resulting in increased adversarial detection capabilities while maintaining overall accuracy during real network attacks in conventional and software-defined networks.

Debicha et al. [43], addressed the vulnerability of machine learning-based NIDS to adversarial attacks, specifically evasion attacks. The research investigates the feasibility of executing such attacks under realistic constraints. The contributions of the study include a detailed analysis of the constraints necessary for generating valid adversarial perturbations while preserving the logic of network attacks. The author introduced a black-box adversarial algorithm capable of developing botnet traffic. The generated adversarial examples can easily evade NIDS without specific knowledge of the target system. The author comprehensively evaluated the results in terms of classification report and detection rate.

Debicha et al. [44], in this study, the authors address the vulnerability of state-of-the-art intrusion detection systems (IDS) based on deep learning to adversarial attacks. They propose a novel approach using transfer learning-based adversarial detectors. The experiments involve implementing existing state-of-the-art IDS models and subjecting them to evasion attacks. The author designed transfer learning-based adversarial detectors, each receiving a subset of information passed through the IDS. Collectively, they demonstrate improved detectability of adversarial traffic in a parallel IDS design. The paper emphasizes the need for sophisticated defence mechanisms, highlighting the shift from simple network protections to advanced IDS systems in the evolving cybersecurity landscape.

Table 1, summarises the state-of-the-art research we have explored in adversarial machine learning and NIDS domain.

Table 1 State-of-the-art Related Research in NIDS and Adversarial Machine Learning

| Study | ML and DL Algorithms | Adversarial Attack Technique | Adversarial Defence Technique | Dataset | Metrics | Key Highlights |
|---|---|---|---|---|---|---|
| Clements et al. [32] (2021) | DL based NIDS, AE, Kitsune | FGSM, JSMA, C&W, ENM | NA | Kitnet Dataset | FPR, FNR, Success Rate, Accuracy | • Investigating system resilience against adversarial attacks. <br> • Attacker had knowledge of the target model for direct adversarial input generation. |
| Usama et al. [34] (2019) | GAN, DNN, SVM, RF, LR, KNN, DT, GB | GAN-based Adversarial Attack | GAN-based Adversarial Defence | KDDCUP-99 | Accuracy, Precision, Recall, F1-Score | • Demonstrated GAN effectiveness in countering adversarial perturbations to enhance NIDS resilience. <br> • Unique method, altered only non-functional features during attack and defence. <br> • Minimizes impact on genuine network operations. |
| Pawlicki et al. [35] (2020) | ANN, RF, SVM, AdaBoost | FGSM, PGD, BIM, C&W | ML based Adversarial Attack Detector | CICIDS-2017 | Accuracy, Precision, Recall, F1-Score | • In-depth exploration and analysis of the impact of adversarial attacks across various ML classifiers in the domain of network security. |
| Guo et al. [36] (2021) | MLP, CNN, SVM, ResNET | BIM | NA | KDDCUP-99, CSE-CIC-IDS2018 | Recall, Confusion Matrix | • ML vulnerabilities to adversarial examples <br> • Proposal of a black-box attack against anomaly network flow detection algorithms <br> • Substitution of a model to create adversarial examples |

| Reference | Models | Attack Method | Defence Method | Dataset | Metrics | Key Contributions |
|---|---|---|---|---|---|---|
| | | | | | | • High probability of bypassing target model detection |
| Alhajjar et al. [37] | SVM, DT, NB, KNN, RF, GB, LR, MLP, LDA, QDA, BG | PSO, GA, GAN | NA | NSL-KDD, UNSW-NB15 | Evasion Rate | • Use of evolutionary computation and deep learning for adversarial example generation. <br> • Evaluation of NSL-KDD and UNSW-NB15 datasets. <br> • Comparison with Monte Carlo simulation. <br> • High misclassification rates in eleven models, show NIDS vulnerability. |
| Sethi et al. [38] | RF, AdaBoost, QDA, GNB, KNN | JSMA | Denoising Autoencoder | NSL-KDD, UNSW-NB15, AWID | Accuracy, False Positive Rate, AUC | • Proposed an IDS design with DRL agents for adaptable attack response. <br> • Integrated ensemble technique with DQN, achieving accuracy-FPR balance. <br> • Implemented fine-grained attack classification post-detection, ensuring high accuracy. <br> • Enhanced robustness against adversarial attacks by proposing DAE integration with reinforcement learning. |
| Zhang et al. [39] | MLP, CNN, C-LSTM, Ensemble | Black-Box Attack Method, PointWise, Opt-Attack, NES, Boundary Attack, HopSkipJump Attack | Ensembling Method, Adversarial, Training, Query Detection. | CICIDS-2017 | Accuracy, Precision, Recall, F1-Score, ASR, MAPE | • Developed TIKI-TAKA framework to assess NN-based NIDS robustness. <br> • Identified vulnerabilities in NIDS against five adversarial attack types. <br> • Proposed defence mechanisms, including model voting and adversarial training. |

| | | | | | | | |
|---|---|---|---|---|---|---|---|
| | | | | | | | • Achieved near 100% intrusion detection rates against most malicious traffic. |
| Han et al. [40] | KitNET, LR, DT, SVM, MLP, IF | Traffic Mutation | Adversarial Feature Reduction | Kitsune Dataset, CICIDS-2017 | Precision, Recall, F1-Score, MER, PDR, DER, MMR | | • Studied gray/black-box traffic-space adversarial attacks in ML-based NIDSs. |
| | | | | | | | • Proposed an attack mutating traffic with limited knowledge, preserving functionality. |
| | | | | | | | • Demonstrated effectiveness against diverse NIDSs and ML/DL models. |
| | | | | | | | • Introduced a defense scheme, reducing evasion rate by >50% in most cases. |
| Maarouf et al. [41] | CNN, DNN, KNN, RNN, C4.5 | DeepFool, PGD, Zoo | NA | SCX-VPN-NON-VPN, NIMS | Accuracy, Precision, Recall, F1-Score | | • Studied ML/DL resilience in encrypted traffic classification. |
| | | | | | | | • Tests C4.5, KNN, ANN, CNN, RNN against evasion attacks. |
| | | | | | | | • Addressed challenges in growing encrypted traffic using AI. |
| | | | | | | | • Deep learning vs. machine learning comparison in adversarial settings. |
| Sarıkaya et al. [42] | RAIDS, LightGBM, k-NN, NN, | WGAN, CTGAN | Adversrial Training | CICIDS 2017, InSDN | Accuracy, Precision, Recall, F1-Score | | • Demonstrated GAN-based attacks on ML-based IDS/IPS systems. |
| | | | | | | | • Trained ML classifier based on autoencoder predictions. |
| | | | | | | | • Introduced RAIDS, a resilient IDS model using autoencoder errors and multiple feature sets. |

| | | | | | | | • Shows RAIDS boosts accuracy by 13.2% and F1-score by 110% against adversarial attacks. |
|---|---|---|---|---|---|---|---|
| Debicha et al. [43] | MLP, RF, KNN | Sign Method | Adversarial Detector | CSE-CIC-IDS2018, CTU-13 | Accuracy, Precision, Recall, F1-Score | | • Explored evasion attacks on ML-based NIDS
• Addressed ML NIDS vulnerability
• Introduced a defence mechanism to safeguard NIDS.
• Assessed the proposed algorithm using realistic botnet traffic for undetected malicious activity. |
| Debicha et al. [44] | DNN | FGSM, JSMA, PGD, DeepFool | Transefer Learning based Adversarial Detector | NSL-KDD, CICIDS 2017, | Accuracy, Precision, Recall, F1-Score Detection Rate | | • Addressed the vulnerability to adversarial attacks in NIDS.
• Introduced transfer learning for robust adversarial detection.
• Evaluates parallel IDS design with strategically placed detectors, showing enhanced detectability. |

## 4. Dataset Description

The CICIDDoS (Canadian Institute for Cybersecurity Distributed Denial-of-Service) dataset is used for experimentation [45]. This dataset is more realistic and represents real physical world attacks compared to older versions of network traffic datasets like KDDCUP-99, NSLKDD and CICIDS-2017 [46]. It includes both benign network traffic and the most recent DDoS attacks. The complete dataset is captured within two days. The first day contains seven attack classes, including PortMap, NetBIOS, LDAP, MSSQL, UDP, UDP-Lag, and SYN. The day two contains twelve classes, namely, SNMP, SSDP, UDP, UDP-Lag, etc. This dataset is publicly available in both pcaps and CSV formats for ML and DL applications. It contains more than 80 features, namely, Flow-ID, Source-IP, Source Port, Destination-IP, Fwd Packet Length Max, Fwd Packet Length Min, etc. These features represent the statistical measurements of network traffic flow logs. However, we have used the random subset of the CICDDoS dataset containing more than 0.25 million samples, 107764 benign samples and 119384 DDoS attack samples for model evaluation purposes.

The dataset has been pre-processed for model training and testing as described in Figure 2 . It has been checked for any null and infinity values and replaced with mean values. The scikit-learn standard scalar function is used to normalize the dataset, as shown in Equation (1). Here, $x$ is the original input data point, $\mu$ represents the mean value, and $\sigma$ represents the standard deviation of the data. For training and testing reasons, the entire dataset has been divided in the ratio of 60:40, with a random state parameter set to 42. Table 2 shows a detailed description of the dataset and its sample counts.

$$x = \frac{x - \mu}{\sigma} \tag{1}$$

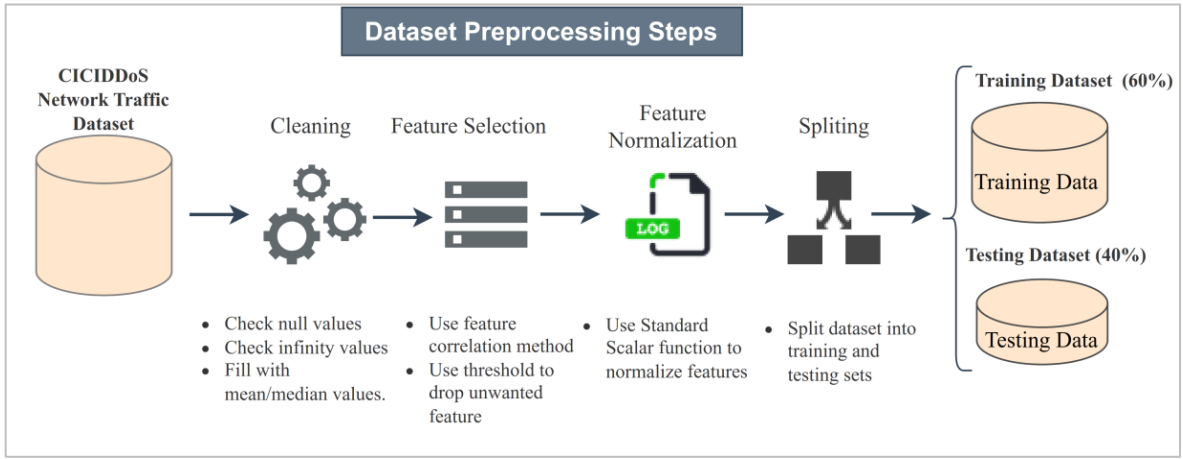

Figure 2 Dataset Pre-processing Steps

Table 2 Dataset Description

| Details | Counts |
|---|---|
| Total dataset samples | 227148 |
| Training samples | 136288 |
| Testing samples | 90860 |
| Training - Testing Split | 40% |
| Random state | 42 |
| Total benign and attack samples | ('BENIGN', 'DDoS') – (107764, 119384) |
| Training benign and attack samples | ('BENIGN', 'DDoS') – (64571, 71717) |
| Testing benign and attack samples | ('BENIGN', 'DDoS') – (43193, 47667) |

## 5. Methodology

This section describes the complete methodology for implementing a black-box adversarial transferability attack on the DL network attack detection model. Our main aim is to explain each detail, starting from model building to attack execution and its evaluation, in simplistic language. So that new researchers can easily understand the technical implementation of the proposed work for future enhancement. The data analyzed, and code generated during the study will be available upon reasonable request. The technical architecture of DL models, FGSM adversarial attack technique, algorithms are as follows.

*5.1. Deep Neural Network*

A deep neural network (DNN) is a type of artificial neural network consisting of one input-layer and one or more hidden layers, followed by the output-layer. All layers consist of multiple neurons, and each layer is interconnected with feed-forward connections, as shown in Figure 3. The activation function between the hidden layers allows DNN to learn complex patterns and hidden representations of the input data. The widely known activation functions are sigmoid, softmax, tanh and relu. The formulation of each activation function is shown in Equation (2) to Equation (4) In our experimentation, we have used the relu activation function between the hidden layers, sigmoid and softmax for the output-layer. The supervised DNN algorithm is used for the - experimentation. Table 3 describes the technical details of hyperparameters of the DL model.

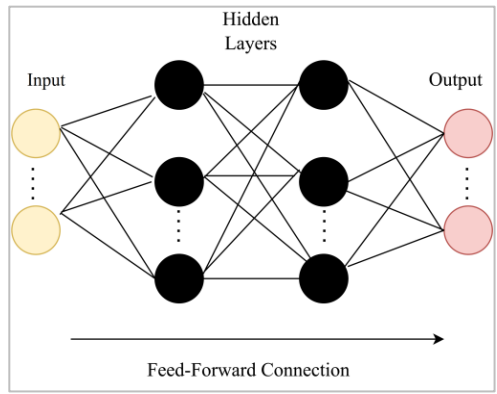

Figure 3 Deep neural network

Table 3 Technical description of the hyperparameter of the deep learning model

| Hyperparameter | Details |
| --- | --- |
| Learning_rate | It is the step-size, used to update the model's parameters during the training process. Choosing an appropriate learning rate is crucial because it affects how quickly the model converges and whether it converges to an optimal solution. |
| Batch_size | It represents the number of training instances (count) in each iteration of the optimization algorithm (such as stochastic gradient descent) before updating the model's parameters. Larger batch sizes can lead to faster training since more examples are processed before each update. However, smaller batch sizes can lead to faster iterations through the dataset, potentially leading to faster convergence. |
| Number of epochs | It represents the number of times the DL algorithm goes through the entire training dataset. In each epoch, the algorithm processes all the training examples once, updating the model's parameters based on the calculated gradients. |
| | It's common practice to monitor the model's performance on a validation set during training. If the validation performance starts to degrade after a certain number of epochs, you can stop training early to prevent overfitting. |
| Activation function | It introduces non-linearity to the algorithm, allowing it to learn complex relationships and predict accurately. The choice of activation function depends on factors such as the network architecture, problem type, and empirical performance on validation data |
| Optimizer | An optimizer is a critical component in training deep learning models. It is an algorithm that adjusts the parameters of the model during the training process to minimize the error (loss). It measures the difference between the prediction and the actual target values. The primary goal of an optimizer is to guide the model towards finding the optimal set of parameters that results in the best performance on the given task. |

$$Sigmoid\ f(x) = \frac{1}{1 + e^{-x}} \tag{2}$$

$$ReLU\ f(x) = max(0, x) \tag{3}$$

$$Softmax\ f(x) = \frac{e^{x_i}}{\sum_{j=1}^{n} e^{x_j}} \tag{4}$$

### 5.1.1. Target Model:

The target system is the DL model, a network attack detection system. The malicious user wants to exploit this system through an adversarial attack. However, the hacker does not have information of the target model and its parameters but has access to the training data. The target system is just like a black-box to the hacker. In real-world situations, the hacker can communicate with the target system only through API to get insights and knowledge about it. The technical architecture of the target used for the experimentation is illustrated in Table 4.

The input shape represents the feature counts in the input training data. The shape of the output layer is set to two for binary classification (benign or attack). The target model architecture comprises two hidden layers, with 50 and 25 neurons in each layer, respectively. The activation functions used are the ReLU between the hidden layers and softmax in the output layer. Adam is used as an optimizer, and the learning rate is set to 1e-3. The loss function is the binary cross entropy for final classification. The validation split parameter is set to 0.2 during the training procedure with a batch size of 4048.

### 5.2. Surrogate Model:

In a real-world situation, the hacker might not have any information about the target system. Creating adversarial examples can be computationally expensive and time-consuming, especially if the hacker does not have direct access to the target model's architecture and parameters. To execute the transferability attack, the attacker must have to create a surrogate model that mimics the behaviour of the target system. Hence, we have created the surrogate DL model with different hyperparameter sets but the same training dataset to approximate or gain insights into the behaviour of the target model.

Three hidden layers are used, with 60, 50, and 30 neurons in each. ReLU activation function is used between the hidden layers and sigmoid in the output layer. Adam is used as the optimizer, and the learning rate is set to 1e-4. The loss function is the binary crossentropy for final classification. The validation split is set to 0.3 during the training procedure with a batch size of 1024, as shown in Table 4.

We have used the less complex architecture for both models because our goal is not to get the optimal results but to explain the working of black-box transferability phenomena in adversarial machine learning. The next step is to generate the adversarial perturbation input with FGSM. The next subsection describes the FGSM technique in detail.

Table 4 Models architecture descriptions

| Hyper-parameter | Surrogate Model | Target Model |
|---|---|---|
| Input shape | train_data.shape | train_data.shape |
| Output shape | 2 | 2 |
| Hidden layers | 3 | 2 |
| Neurons | 60,50,30,2 | 50,25,2 |
| Activation function – hidden layer | ReLU | ReLU |
| Activation function – output layer | sigmoid | softmax |
| Optimizer | Adam | Adam |
| Learting rate | 0.0001 (1e-4) | 0.001 (1e-3) |
| Loss | binary_crossentropy | binary_crossentropy |
| Validation split | 0.3 | 0.2 |
| Batch size | 1024 | 4048 |
| Trainable params # | 3322 | 1,982 |

## 5.3. Fast Gradient Sign Method (FGSM)

FGSM attack is an untargeted evasion adversarial attack that takes place during the inference or testing time. The main goal of FGSM is to create adversarial perturbation examples by adding tiny noise into the model input. These adversarial examples are crafted very carefully and can easily deceive the model with a high-confidence score. FGSM is a gradient-based approach that optimizes the $L_P$ norm by taking one step in the opposite direction of the gradient to each element of input vector $x$. The optimal max-norm constrained perturbation is formulated as in Equation (5) – (7). The detailed step-by-step technical description of adversarial perturbation input generation is given as follows:

**Gradient input**: In order to generate an adversarial perturbation, the gradient of the loss function must be determined with respect to the input $x$.

$$gradient \;=\; \nabla xJ(\Theta, x, y) \qquad (5)$$

**Gradient sign:** then gradient sign is calculated for each feature in the input sample *x* to know in which direction the loss function would increase if the perturbation is added to the feature.

$$gradient\_sign \;=\; sign(\nabla xJ(\Theta, x, y)) \qquad (6)$$

**Add perturbation:** add the perturbation value $\epsilon$ (small noise) into the model input to generate the adversarial perturbation example.

$$x_{adv} \;=\; x \;+\; gradient\_sign \qquad \textbf{(7)}$$

**Output**: the resulting output $x_{adv}$ is the adversarial perturbation input that fools the model into producing the incorrect result.

Figure 4, demographically presents the adversarial perturbation examples generated using the surrogate model and testing dataset. Since FGSM is a white-box method, it requires a classifier to generate adversarial perturbation examples. The surrogate model uses a classifier that mimics the target classifier that needs to be compromised through adversarial attack. The method generates a list of adversarial perturbations for different epsilon values ranging from 0.0001 to 0.0009. we did this to comprehensively evaluate the NIDS model for both white-box and black-box transferability attacks.

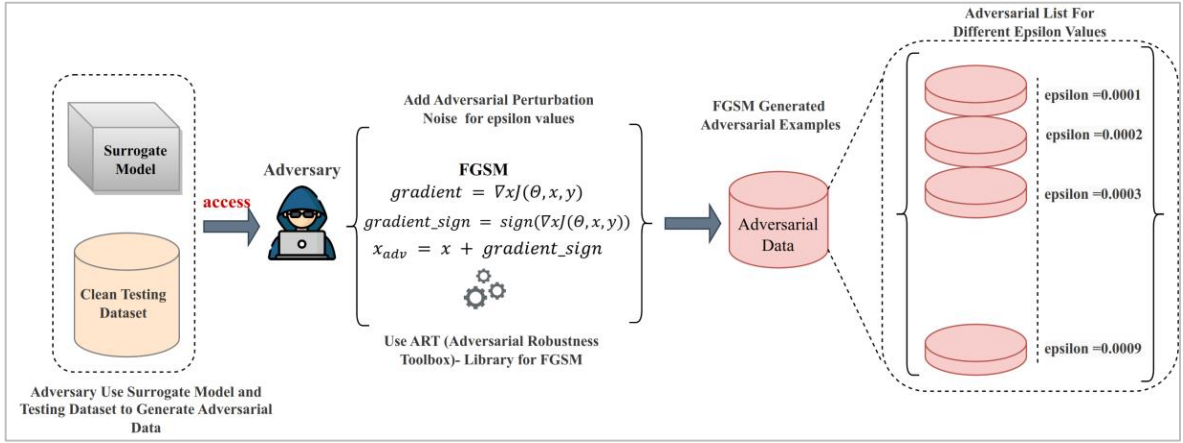

Figure 4 Generation of Adversarial Perturbation Examples using FGSM

### 5.4. Conceptual architecture and algorithms

The conceptual description of the proposed algorithm is illustrated in Figure 5. As shown, both the admin and hacker can access to the entire training data. The goal is to deceive the target model to which the hacker can communicate only through API. We have tested the proposed algorithm in two ways: the first one is a white box adversarial attack, and the second is a black box adversarial transferability attack. The white-box adversarial attack required full access to the model. In our case, the adversarial perturbation is generated based on the surrogate model. Hence, deceiving the surrogate model with the adversarial perturbation inputs is the white-box adversarial attack.

On the other hand, in the black-box adversarial transferability attack, the same adversarial perturbation input is used to execute a transferability adversarial attack on the targeted system. The evaluation of both models is done before and after the attack implementation. It is concluded the white-box attack has a severe effect compared to the black-box transferability attack in terms of reducing the confidence score of the models.

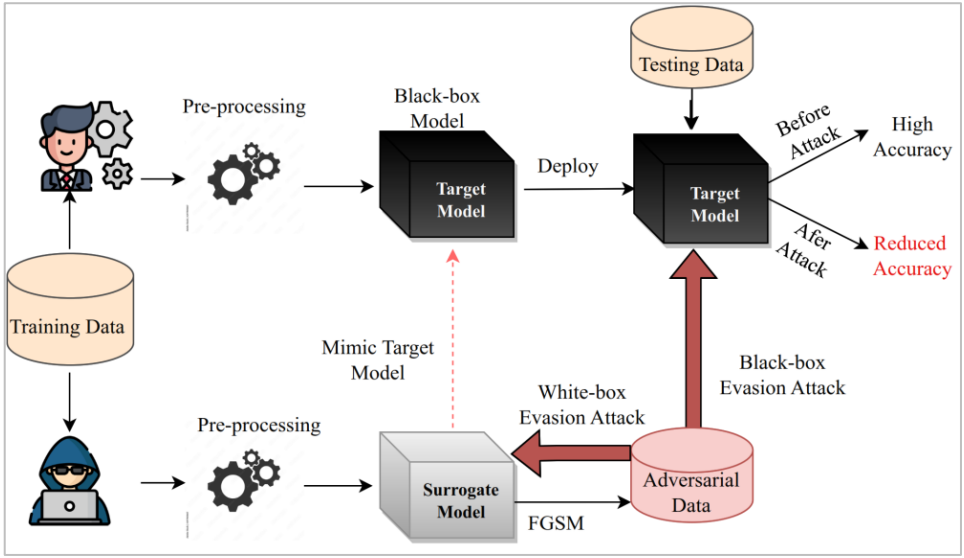

Figure 5 Conceptual diagram of the proposed research study

**Algorithm 1** and **Algorithm 2** describe the detailed step-by-step procedure of technical implementation of the adversarial perturbation inputs generation and attack execution. In **Algorithm 1**, we have explained how the adversarial perturbation examples are generated using the FGSM [47]. The KerasClassifier method is used in which the surrogate model is passed as a parameter. The next parameter we address is the min and max values within the training dataset. These values serve as reference points to ensure that the generated perturbations remain within a predefined range. This "clipping" process prevents the perturbations from venturing beyond acceptable bounds. Then, we define the "fgsm_attack" function, which accepts two parameters: the classifier (the model under consideration) and the epsilon value. This function is designed to generate and provide the adversarial perturbation input for a specific epsilon value. The outcome of this function is a perturbed input that is carefully crafted to deceive the model. We applied the "fgsm_attack" function across various epsilon values, which span the range from 0.0001 to 0.0009. By executing the function for each epsilon value, we collect a series of adversarial perturbation inputs.

Algorithm 2 encompasses two sections; the first section is the implementation of a white-box adversarial attack, while the second focuses on executing a black-box attack based on the list of previously generated adversarial perturbations obtained from Algorithm 1.

The "model_prediction" function is defined to provide a classification report for the input data. The next step evaluates both the surrogate model and the target model using clean test data. This initial evaluation is conducted in the absence of any adversarial attack, allowing us to establish a baseline understanding of their performance. The third step encompasses the implementation of the white-box attack. The adversarial perturbations are utilized to evaluate the surrogate model over the different epsilon values. In the fourth step, the black-box adversarial transferability attack is implemented. During this phase, the previously generated adversarial perturbation examples are deployed in the context of the target system during inference. This enables us to explore the transferability of these perturbations from the surrogate model to the target model, gauging the extent to which adversarial attacks can cross over between different models.

---

**Algorithm 1** : Adversarial perturbation generation with FGSM and surrogate model

**Input**: surrogate_model, test_data, list_of_epsilon

**Output**: list of adversarial perturbation input

**Steps:**

**Step 1: Build keras classifier with the surrogate model and clip the values between max and min of test data**.

    classifier = KerasClassifier(model=surrogate_model, clip_values = (np.min(test_data),
    np.max(test_data)), use_logits = False)

**Step 2: Define a function to generate adversarial perturbation with epsilon value.**

    **define** fgsm_attack(classifier, test_data, epsilon):
        fgsm = FastGradientMethod(classifier, eps= epsilon)
        adversarial_input = fgsm.generate(test_data)
   **return** adversarial_input

**Step 3: Call the function in a loop to generate adversarial input for multiple values of epsilon**

    adversarial_input_list =  []
    **for** epsilon **in** list_of_epsilon:
       adversarial_input = fgsm_attack (classifier, test_data, epsilon)
       adv_list.append(adversarial_input)
   **end for**

---

**Algorithm 2** – White-box and Black-box transferability adversarial attack implementation with FGSM

**Input**: surrogate_model, target_model, test_data, adversarial_input_list

**Output**: classification_report

---

*Preprint Submitted to Computer & Security*                                                                                            April 12, 2024

**Steps:**

**Step 1: Define the prediction function for model evaluation**
```
define model_prediction(model, input_data):
    prediction = model.predict(input_data)
    classification_report = (actual_input, prediction)
    return classification_report
```

**Step 2: Evaluation of both model before adversarial attack**
```
Classification_report_surrogate = model_prediction(surrogate_model, test_data)
Classification_report_target = model.predict(target_model, test_data)
```

**Step 3: Evaluate the white-box attack on the surrogate model by passing an adversarial input data list for different epsilon.**

```
classification_report_surrogate=[]
for adversarial_input in adversarial_input_list:
    classification_report = model_prediction(surrogate_model, adversarial_input)
    classification_report_surrogate.appen(classification_report)
    print(classification_report)
end for
```

**Step 4: Evaluate the black-box attack transferability attack on the model by passing an adversarial input data list for different epsilon.**

```
classification_report_target=[]
for adversarial_input in adversarial_input_list:
    classification_report = model_prediction(target_model, adversarial_input)
    classification_report_target.append(classification_report)
    print(classification_report)
end for
```

## 6. Experimental Setup and Supportive Libraries

This section describes all supportive libraries used for proposed research work, followed by system configuration. It also describes the testbed environment used for the experiments followed by the model evaluation metrics such accuracy, precision, recall.

*6.1. Setup and Supportive Libraries*

The experiment is conducted on Google Colab, a cloud-based platform with free GPU and TPU support for ML and DL applications. It is integrated with Google Drive, hence further simplifying the procedure of saving, loading and downloading the data files. It supports a wide range of popular libraries such as keras for DL model building, pandas for dataset manipulation and analysis, scikit-learn for dataset preprocessing and normalization, matplotlib for plotting the graph and data visualization and much more as shown in Table 5. This eliminates any hardware and software installation in the local machine and compatibility issues. It is a time and cost-saving platform for ML and DL. The ACER laptop equipped with Window-11 Operating System, Core i7 processor, 500 SSD, and 16 GB DR4L RAM is used for the experimentation, as described in Table 6.

Table 5 Supportive Libraries for Experimentation

| Experimentation | |
|---|---|
| Platform | Description |
| Google Colab | Free cloud platform for deep learning experiments. Comes with GPUs and TPUs. Works well with Google Drive and Jupyter Notebook. Easily run Python code interactively. |
| **Required Installation for Adversarial Machine Learning** | |
| Adversarial Robustness Toolbox (ART) | Toolkit designed for evaluating and enhancing the robustness of deep learning models against adversarial attacks. |
| **Pre Installed Libraries in Google Colab Platform** | |
| Keras | Library for a high-level neural network API, Used to create DL model in the proposed research work. |
| Pandas | Dataset analysis library utilized for loading and analyzing the dataset in the development of NIDS. |
| NumPy | The library supports large multi-dimensional arrays and mathematical functions. It is used for converting data into NumPy arrays during NIDS model training and testing. |
| Matplotlib | Matplotlib Library, used for visual representation. Instrumental in visually presenting NIDS model training plots, enhancing the interpretability of experimental outcomes. |
| Scikit-Learn | Machine Learning library offers various evaluation metrics (accuracy, precision, recall, f1-score, AUC-ROC) for NIDS model evaluation. |

Table 6 Experimentation environment

| Name | Details |
|---|---|
| Operating System | Window-11 |
| Processor (CPU) | Core(TM) i7-10870H (2020) |
| Base Clock Performance | 2.20 GHz |
| Turbo Clock Performance | 5.00 GHz |
| SSD | 500 |
| RAM | 16 GB-DDR4L |
| GPU Support | NVIDIA-GEFORCE GTX |

## 6.2. Google Colab Testbed Environment for Experimentation

Figure 6, illustrates the procedural flow diagram of our Google Colab testbed environment utilized for experiments. The initial step involves the creation of a directory folder, which is structured to isolate files for results, saved models, logs, etc. Subsequently, Google Drive is connected to the Colab platform for experimentation. It is a crucial step to import and export of files between Google Drive and Colab. The Colab platform comes equipped with pre-installed libraries such as pandas, keras, numpy, etc., which are essential for ML and DL research. The installation of the ART library is required for implementing adversarial attacks method. Following this setup, we generate a Jupyter file for our Python code. The effectiveness of this configuration becomes clear as it allows for the seamless saving of experiment results directly to Google Drive for future reference, as shown in Figure 6.

In conclusion, Google Colab is an excellent platform for research in ML and DL. It offers support for GPU and TPU, which helps train DL models quickly. Colab also comes with pre-installed libraries needed for experiments, making code running environment easier. It is user-friendly, especially for setting up the environment. We recommend new researchers to use Google Colab for their ML and DL research studies.

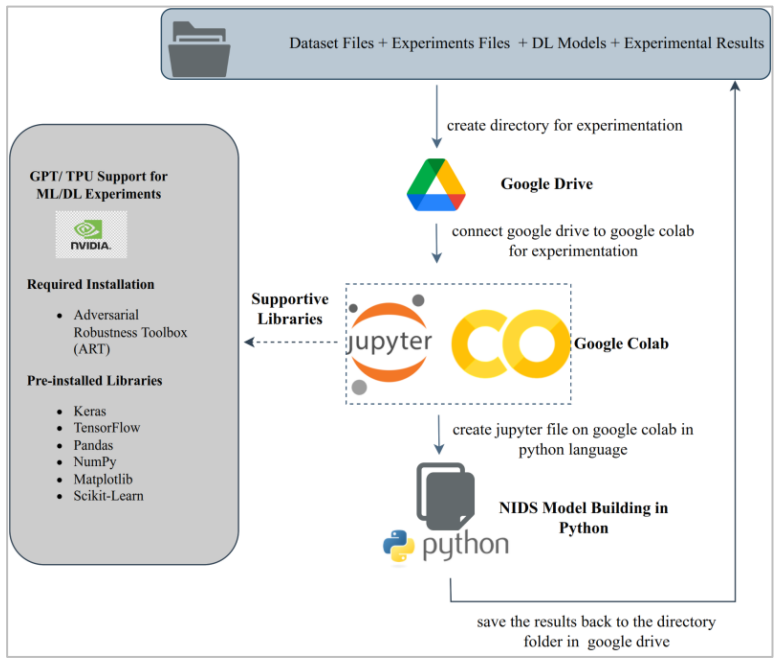

Figure 6 Testbed Architecture for Experimentation using Google Colab

## 6.3. Performance Metrics:

The performance of both the models is evaluated based on accuracy, precision, recall, f1-score and ROC AUC curve. The brief description of each is as follows:

Accuracy: It represents the correct classification of the samples by the model out of all the available samples. It gives the overall assessment of the model.

Precision: It is the proportion of the actual true positives predicted by the model out of all positive predicted samples. High precision indicates low false positives predicted by the model.

Recall: It measures the proportion of true positive predictions among all actual positive samples. It is important when the focus is to reduce false negatives.

F1-score: It is a harmonic mean of precision and recall. It is a good measure when the dataset is imbalanced.

Area Under the ROC Curve (AUC-ROC): This represents the area under the Receiver Operating Characteristic curve, which plots the true positive rate against the false positive rate. It measures the model's ability to discriminate between classes.

Equations (8) - (11) show the formulation of performance metrics in terms of true positive (TP), false positive (FP), true negative (TN) and false negatives (FN).

$$ACC = \frac{TP + TN}{TP + TN + FP + FN} \tag{8}$$

$$Precision\ (P) = \frac{TP}{TP + FP} \tag{9}$$

$$Recall\ (R) = \frac{TP}{TP + FN} \tag{10}$$

$$F - Score\ = \frac{2 \times R \times P}{R + P} \tag{11}$$

## 7. Results and Discussion

In this section, we discussed the obtained results from the surrogate and target model in terms of classification report and confusion matrix. We have done a comprehensive analysis and discussion on the results, starting from the model's training process to when subjected to white-box and black-box transferability attacks.

*7.1. Surrogate and Target Model Performance*

We kept the simple architecture of both the surrogate and the target DL model. It is because our objective is not to focus on optimizing the outcomes but rather to conduct empirical testing of the black-box transferability phenomenon.

The surrogate and target models are trained on the same dataset but with different hyperparameter sets, as previously discussed. We have used ModelCheckpoint, a powerful tool to save the best model automatically during the training and validation procedure. It avoids overfitting and ensures that the most optimal weights are stored to get the best model. Both models are trained for binary classification of benign and attack classes on the CICDDoS-2019 dataset. Table 7 demonstrates the training and validation results of the surrogate model at regular intervals of the epochs. The training accuracy is slightly aligned with the validation accuracy, indicating that the model does not overfit the training data and can generalize well on the unseen dataset. The surrogate model is trained for 50 epochs and yielding a remarkable accuracy of 98.97% on the training data and 98.98% on the validation data. These findings validate the results of the surrogate model. Furthermore, the loss of the model has also been notably reduced for both the training and validation stages, 0.0330 and 0.0323, respectively. The identical training approach is employed to train the target system; however, the target model underwent 60 epochs. The obtained peak training and validation accuracy are 97.49% and 97.54%, respectively. Notably, the model's performance did not demonstrate overfitting tendencies and achieved satisfactory results, as shown in Table 8.

Table 7 Surrogate model training and validation results

| No. of Epochs | Values |
|---|---|
| Epoch-1/50 | loss:- 0.6467 - accuracy: 0.6781 – val-loss: 0.5624 - val-accuracy: 0.7603 |
| Epoch-5/50 | loss:- 0.2469 - accuracy: 0.9218 – val-loss: 0.2160 – val-accuracy: 0.9354 |
| Epoch-10/50 | loss:- 0.1012 - accuracy: 0.9721 – val-loss: 0.0988 – val-accuracy: 0.9727 |
| Epoch-15/50 | loss:- 0.0761 - accuracy: 0.9762 – val-loss: 0.0806 – val-accuracy: 0.9750 |
| Epoch-20/50 | loss:- 0.0663 - accuracy: 0.9776 – val-loss: 0.0683 – val-accuracy: 0.9780 |
| Epoch-25/50 | loss:- 0.0589 - accuracy: 0.9796 – val-loss: 0.0626 – val-accuracy: 0.9778 |
| Epoch-30/50 | loss:- 0.0520 - accuracy: 0.9813 – val-loss: 0.0556 – val-accuracy: 0.9834 |
| Epoch-35/50 | loss: -0.0463 - accuracy: 0.9837 – val-loss: 0.0471 – val-accuracy: 0.9849 |
| Epoch-40/50 | loss: -0.0396 - accuracy: 0.9862 – val-loss: 0.0410 – val-accuracy: 0.9870 |
| Epoch-45/50 | loss:- 0.0361 - accuracy: 0.9880 – val-loss: 0.0396 – val-accuracy: 0.9904 |
| Epoch-50/50 | loss: -0.0330 - accuracy: 0.9897 – val-loss: 0.0323 – val-accuracy: 0.9898 |

Table 8 Target model training and validation results

| No. of Epochs | Values |
|---|---|
| Epoch 1/60 | loss:- 0.6849 - accuracy: 0.5622 - val_loss: 0.6710 – val-accuracy: 0.6072 |
| Epoch 5/60 | loss:- 0.5681 - accuracy: 0.7220 - val_loss: 0.5485 – val-accuracy: 0.7474 |
| Epoch 10/60 | loss:- 0.4363 - accuracy: 0.8201 - val_loss: 0.4287 – val-accuracy: 0.8257 |
| Epoch 15/60 | loss:- 0.3423 - accuracy: 0.8554 - val_loss: 0.3345 – val-accuracy: 0.8623 |
| Epoch 20/60 | loss:- 0.2490 - accuracy: 0.9359 - val_loss: 0.2439 – val-accuracy: 0.9389 |
| Epoch 25/60 | loss:- 0.1864 - accuracy: 0.9592 - val_loss: 0.1847 – val-accuracy: 0.9582 |
| Epoch 30/60 | loss:- 0.1493 - accuracy: 0.9657 - val_loss: 0.1500 – val-accuracy: 0.9650 |
| Epoch 35/60 | loss:- 0.1276 - accuracy: 0.9681 - val_loss: 0.1298 – val-accuracy: 0.9684 |
| Epoch 40/60 | loss:- 0.1140 - accuracy: 0.9698 - val_loss: 0.1166 – val-accuracy: 0.9702 |

| Epoch 45/60 | loss:- 0.1045 - accuracy: 0.9710 - val_loss: 0.1076 – val-accuracy: 0.9707 |
| Epoch 50/60 | loss:- 0.0977 - accuracy: 0.9723 - val_loss: 0.1014 – val-accuracy: 0.9714 |
| Epoch 55/60 | loss:- 0.0934 - accuracy: 0.9735 - val_loss: 0.0969 – val-accuracy: 0.9749 |
| Epoch 60/60 | loss: -0.0890 - accuracy: 0.9749 - val_loss: 0.0920 – val-accuracy: 0.9754 |

Figure 7 depicts a visualization of the loss and accuracy plots of surrogate and target models. It indicates how the model loss reduces and accuracy increases as the number of epochs rises for both the surrogate and target models. A smooth curve implies that the model's performance is evolving in a relatively steady manner without abrupt fluctuations or instability. This can suggest that the model is learning effectively and converging towards an optimal solution. We evaluated the models under three situations. The first is before the adversarial attack. It describes the initial state of the model where no intentional modifications, perturbations, or attacks have been applied. The model is evaluated on the clean test dataset. The second is the under-white box adversarial attack. In this case, the adversarial perturbation inputs are applied to the surrogate model to deceive it. The third scenario involves a black box attack conducted on the target model using adversarial perturbation examples generated using the surrogate model.

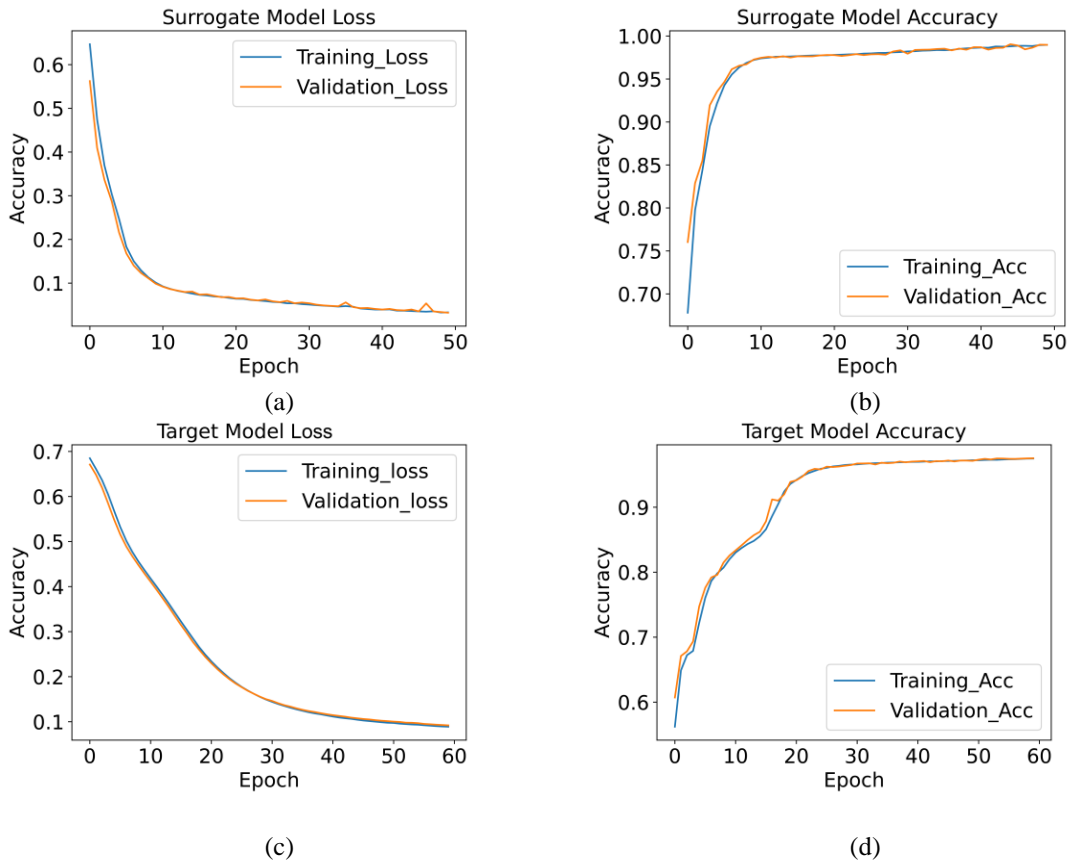

(a)      (b)

(c)      (d)

Figure 7 (a) Surrogate Model Loss (b) Surrogate Model Accuracy (c) Target Model Loss (d) Target Model Accuracy

Table 9, Figure 8, shows the first scenario in which the results are classified in the classification report. The models are evaluated on the clean testing dataset (without adversarial perturbation). Both models achieved excellent results in terms of accuracy, precision, recall and f1-score. The obtained accuracy and f1-score of the surrogate model are 99.05%, and 99.01%, respectively.

Table 10, Table 11, Figure 9 and Figure 10 are showing the confusiom matrix of both models. Table 10, highlights the confusion matrix with a false positive and false negative scenario of 745 and 155, respectively. In the case of the target model, accuracy and f1-score are 97.57% and 97.57%, with false positive and false negative scenarios of 1723 and 482, as shown in Table 10, and Table 11. The surrogate model has a relatively low number of false positives and false negatives. Keep in mind that a high number of false positives indicates that the model is incorrectly classifying instances as positive when they should be negative, and a high number of false negatives indicates that the model is incorrectly classifying instances as negative when they should be positive. The goal is to strike a balance between these two types of errors while maximizing accuracy, precision, recall, and F1-score.

Table 9  Classification report before the attack

| Model | Accuracy % | Precision % | Recall % | F1-Score % |
|---|---|---|---|---|
| Surrogate Model | 99.05 | 98.98 | 99.01 | 99.01 |
| Target Model | 97.57 | 97.61 | 97.57 | 97.57 |

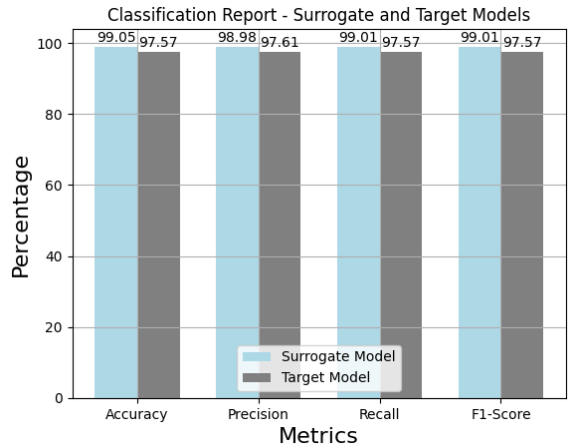

Figure 8 Classification Report of Surrogate and Target Model Before Attack

Table 10  Confusion matrics before attack surrogate model

| LB | Predicted |
|---|---|

|        |        | Benign | Attack |
|--------|--------|--------|--------|
| Actual | Benign | 42448  | 745    |
|        | Attack | 155    | 47512  |

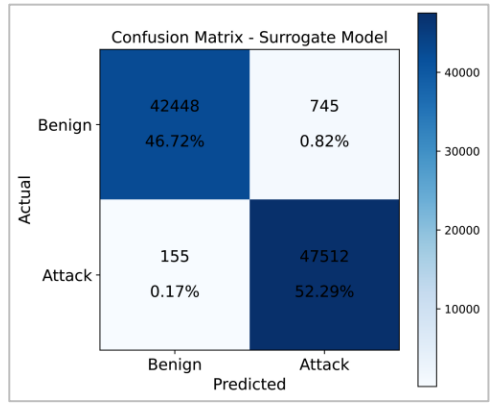

Figure 9 Confusion Matrix Before Attack Surrogate Model

Table 11 Confusion matrices before attack Target Model

| LB     |        | Predicted |        |
|--------|--------|-----------|--------|
|        |        | Benign    | Attack |
| Actual | Benign | 41470     | 1723   |
|        | Attack | 482       | 47185  |

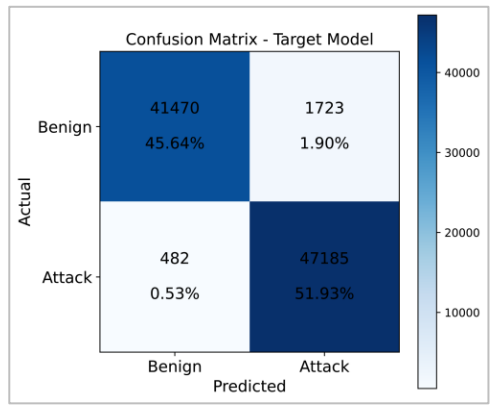

Figure 10 Confusion Matrix Before Attack Target Model

## 7.2. White-box Attack on the Surrogate Model with FGSM

As already discussed, a white-box adversarial attack occurs when the attacker has access to the internal details of the model that they intend to compromise. And based on these details the perturbation examples are generated to deceive the same model (surrogate model). Table 12 demonstrates the classification result under the adversarial attack condition for various epsilon values ranging from 0.0001 to 0.0009. Epsilon is a parameter that determines the magnitude of the perturbations added to the input data. It controls how much the input data is modified while still being perceivable as the original data. The results indicate that as the epsilon value increases (i.e., larger perturbations are applied), the overall classification performance decreases. In other words, the model becomes more susceptible to misclassification as the perturbations become stronger. This observation aligns with the general behaviour of adversarial attacks, where larger perturbations are more likely to cause misclassifications because they push the model further away from its decision boundaries. The accuracy reduced from 99.05% to 23.85%. And the f1-score reduced from 99.01% to 23.00%.

Table 12 Classification report after the white-box attack

| Epsilon value | Accuracy % | Precision % | Recall % | F1-score % |
|---|---|---|---|---|
| 0.0001 | 86.14 | 88.88 | 86.14 | 86.00 |
| 0.0002 | 67.21 | 70.49 | 67.21 | 66.37 |
| 0.0003 | 56.65 | 58.36 | 56.65 | 55.68 |
| 0.0004 | 43.54 | 43.64 | 43.54 | 41.86 |
| 0.0005 | 35.95 | 35.45 | 35.95 | 34.75 |
| 0.0006 | 31.03 | 30.35 | 31.03 | 30.03 |
| 0.0007 | 28.02 | 27.18 | 28.02 | 27.04 |
| 0.0008 | 25.79 | 24.91 | 25.79 | 24.87 |
| 0.0009 | 23.85 | 22.96 | 23.85 | 23.00 |

## 7.3. Black-box attack on the surrogate model with FGSM

The third case is the implementation of the white box adversarial attack on the target model. The adversarial perturbation generated through the surrogate model is applied to the target model in the inference phase. As depicted in the Table 13, the results are classified for different epsilon values. The outcomes exhibit a similar impact as observed in the white-box attack but with a relatively reduced rate. Based on the results, it can be concluded that white-box attacks tend to have a greater magnitude or strength compared to black-box attacks. This suggests that the attacker's knowledge of the internal model details in a white-box attack enables them to craft more effective perturbations, leading to more substantial changes in the model's outputs. The accuracy reduced from 97.02% to 61.15%. And the f1-score reduced from 97.01 % to 59.94%. The classification results of both adversarial attacks are visually represented using different epsilon values. The plot for the white-box attack appears smoother when compared to the plot for the black-box attack across various epsilon values in Figure 11.

Table 13 Classification report after the black-box attack

| Epsilon value | Accuracy % | Precision % | Recall % | F1-score % |
| --- | --- | --- | --- | --- |
| 0.0001 | 97.02 | 97.12 | 96.94 | 97.01 |
| 0.0002 | 96.28 | 96.34 | 96.21 | 96.26 |
| 0.0003 | 95.27 | 95.28 | 95.24 | 95.25 |
| 0.0004 | 93.62 | 93.66 | 93.64 | 93.65 |
| 0.0005 | 91.47 | 91.45 | 91.55 | 91.47 |
| 0.0006 | 79.96 | 81.66 | 80.52 | 79.85 |
| 0.0007 | 76.78 | 7870 | 77.39 | 76.61 |
| 0.0008 | 71.42 | 73.84 | 72.14 | 71.06 |
| 0.0009 | 61.15 | 64.39 | 62.15 | 59.94 |

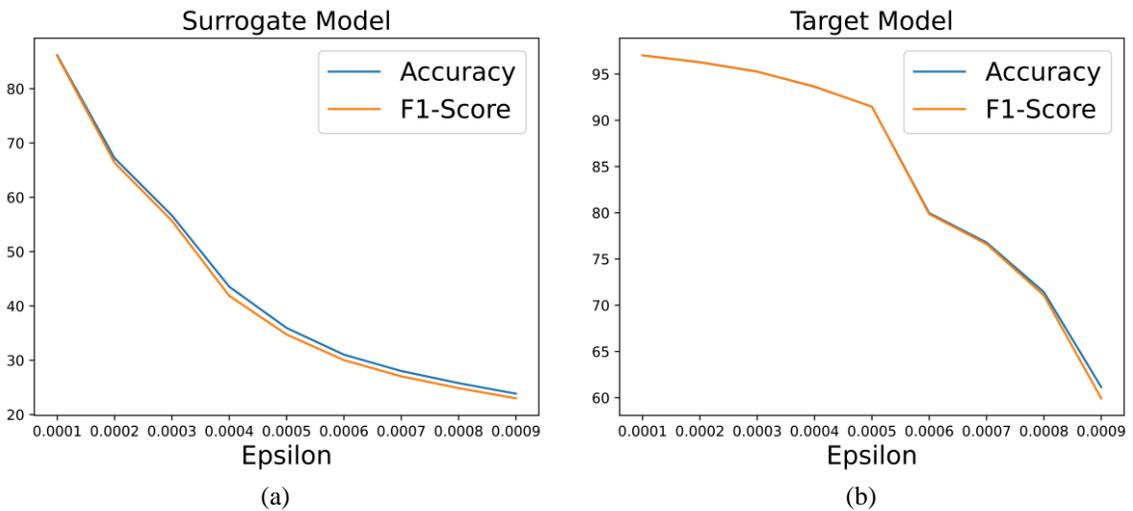

Figure 11. (a) Surrogate model accuracy and f1-score (b) Target model accuracy and f1-score

## 7.4. Discussion

The evaluation of model performance unfolds in three distinct phases. These phases include the initial state without any adversarial attack, a white-box attack on the surrogate model, and a black-box transferability attack on the target model.

No Adversarial Attack: In the absence of any adversarial perturbations, both the surrogate and target models exhibited optimal performance during the initial phase. The classification report and confusion matrices revealed high accuracy, precision, recall, and F1 scores (~99 %). These results indicate that both models are robust when evaluated with clean test data.

White-Box Attack on Surrogate Model: The second phase introduced a white-box adversarial attack on the surrogate model using the FGSM. This scenario represents an adversary who has full knowledge of the surrogate model, such as its DL architecture, gradients, loss function, etc. These details are used by the adversary to generate adversarial perturbation to attack or deceive the surrogate model. As the epsilon value increased from 0.0001 to 0.0009 in a white-box adversarial attack, the surrogate model experienced a substantial decline in accuracy, dropping from 86.14% to 23.85%. The corresponding reductions in precision, recall and f1-score are 88.88% to 22.96%, 86.14% to 23.85%, and 86.00% to 23.00%, respectively. It shows the model's vulnerability to stronger adversarial perturbations.

Black-Box Transferability Attack on Target Model: The third phase involved a black-box transferability attack on the target model. Here, adversarial perturbations generated using the surrogate model are applied to the target model. The impact of the transferability attack was less severe compared to the white-box scenario. The results revealed a reduction in the classification report of the target model. As the epsilon value increased from 0.0001 to 0.0009 in a black-box transferability attack, the target model witnessed a notable decline in performance. The accuracy dropped from 97.02% to 61.15%, with reductions in precision (97.12% to 64.39%), recall (96.94% to 62.15%), and F1-score (97.01% to 59.94%).

In summary, a white-box adversarial attack has a severe attack magnitude compared to a black-box adversarial attack. However, the black box scenario represents more realistic adversarial attacks. As in the real physical world, the adversary has no knowledge of the targeted system that is to be attacked or compromised.

In our previous research [12], we have shown the real-world implementation of white-box adversarial attacks within network packet scenarios. This work illustrates how adversaries can extract and manipulate packet-level information within a Wide Area Network (WAN). We recommend this research work for newcomers who seek a deeper understanding of adversarial attacks in real-world scenarios. It would also provide the practical applicability of the proposed research work. As discussed, this study [12] is for white-box adversarial attack, but the concept is also applicable to black-box and adversarial transferability attacks. We have shown only conceptual idea for black-box adversarial transferability in [12], but this proposed research is an empirical enhancement of black-box adversarial transferability attack.

In addition, further research could explore robust model architectures, training techniques, and defence mechanisms to mitigate the impact of both white-box and black-box adversarial attacks. Additionally, evaluating model performance under various attack scenarios helps understand and address vulnerabilities and resilient deep-learning models in real-world applications.

## 8. Limitations and future directions

This section highlights some limitations of the research work and the future scope for the new researcher in the area of adversarial machine learning. Following are the details pointing out limitations and future scope.

- In this study, our emphasis has been on detecting adversarial attacks from the network security perspective. However, it is important to note that other domains, including Computer Vision, Natural Language Processing, Medical Imaging, and Finance, offer promising opportunities for further exploration in adversarial attack detection.

- This research is constrained by its exclusive focus on using an FGSM attack strategy. Nonetheless, there exist other adversarial attacks like JSMA [48], PGD [49], and C&W [50] to form adversarial perturbation inputs to deceive the model. The future scope of this work may include attacks, such as poisoning attacks, model extraction, and inference attacks, which could serve as promising directions for future extensions.

- The proposed research did not explore the defence method against the adversarial tactics. Hence, future direction is the exploration of different defence strategies, such as adversarial training and filtering methods, to increase the resilience of the DL model. Additionally, ensemble methods can improve the

- robustness of the DL model. The significant benefits of ensemble machine learning involve improved accuracy, reduced overfitting, robustness, generalization, capacity, etc.
- In most of the defence methods, the robustness of the model increases during the training phase. But the same can also be applied during the testing or the inference phase. The strategies to remove the noise or filtering methods can be used for the adversarially generated perturbation inputs.
- The other potential future directions involve exploring the transferability concept into adversarial defence strategies across different models. For example, if one adversarial defence is applicable to one model, will it be equally beneficial for another model?
- The proposed research work is implemented in the testbed development settings. However, the real-time development and deployment of the proposed concept may introduce new challenges to explore.
- The study can also be extended with different intrusion detection datasets and DL architectures to explore the phenomena.

## 9. Conclusion

In conclusion, this research assesses the vulnerability of deep learning models against adversarial attacks in the context of a cyber attack detection system. Our research introduces a novel aspect by implementing black-box adversarial transferability phenomena through surrogate and target models [51]. It says that adversarial perturbation examples generated by one DL model could have a similar impact on the other model even with the different architecture and parameter settings. We have used the FGSM to create adversarial perturbation examples using the latest CICDDoS-2019 dataset. The comprehensive evaluation is done under the three cases, namely, without attack, after the white-box attack on the surrogate model and the black-box transferability attack on the target model. The results obtained in terms of the classification report and confusion matrix offer valuable insights. Notably, we observed that white-box attacks tend to have more severe effects when compared to the black-box counterparts across varying epsilon values (0.0001 to 0.0009). However, the black-box transferability attacks mirror real-world conditions more closely. As external attackers (adversaries) often operate without access to detailed targeted system architecture.

In Addition, we have also added detailed taxonomical information on adversarial attacks with multiple dimensions. This information will help new researchers to explore different domains of research in adversarial machine learning. This insight is crucial for the development of robust defences that can effectively mitigate the impact of adversarial attacks in the real-world adversarial cyber threat landscape.

## **Declarations:**

**Ethical Approval**: Not Applicable

**Consent to participate:**  Not Applicable

**Consent for publication**: Not Applicable

**Authors' contributions**: Author-1 has experimented and written the complete manuscript. Author-2 has reviewed the manuscript and suggested modifications and changes for further improvements.

**Funding:** Not Applicable

**Conflicts of Interest:** The authors declare no conflict of interest.

**Data Availability Statement**: The datasets analyzed during the current study are available online. https://www.unb.ca/cic/datasets/ddos-2019.html

**Code Availability Statement:** The code generated during and/or analyzed during the current study will be available at reasonable request.